\documentclass[twocolumn]{aastex7}
\usepackage{hyperref}
\usepackage{tabularx}
\usepackage{amsmath}
\usepackage{graphicx}
\usepackage{booktabs}
\usepackage{multirow}
\usepackage{tablefootnote}
\let\tablenum\relax 
\usepackage{siunitx}
\usepackage{needspace}  
\usepackage{float} 
\usepackage{placeins}

\newcommand\RE{$R_{\Earth}$}

\newcommand\ME{$M_{\Earth}$}

\shortauthors{Distler and Becker}
\begin{document}

\title{Peas and USPs: Can Stellar Spindown and Peas in a Pod Replicate Ultra-Short-Period Planet Characteristics?}

\author[orcid=0009-0006-4294-6760]{Adam Distler}
\altaffiliation{NSF Graduate Research Fellow}
\affiliation{Center for Astrophysics | Harvard and Smithsonian, 60 Garden Street, Cambridge, MA 02138, USA}
\email{adam.distler@cfa.harvard.edu} 
\affiliation{Department of Astronomy, University of Wisconsin-Madison, 475 N.~Charter St., Madison, WI 53706, USA}
\affiliation{Wisconsin Center for Origins Research, University of Wisconsin--Madison, 475 N Charter St, Madison, WI 53706, USA}
\email[show]{adamdistler@cfa.harvard.edu}

\author[orcid=0000-0002-7733-4522]{Juliette Becker}
\affiliation{Department of Astronomy, University of Wisconsin-Madison, 475 N.~Charter St., Madison, WI 53706, USA}
\affiliation{Wisconsin Center for Origins Research, University of Wisconsin--Madison, 475 N Charter St, Madison, WI 53706, USA}
\email{juliette.becker@wisc.edu}

\begin{abstract}
Peas-in-a-Pod (PIAP) systems have been shown to be common across exoplanet systems, with regular planet spacings and similar planet sizes. In contrast, ultra-short-period planets have displayed distinct differences from PIAP systems, including higher mutual inclinations, ages, and planet sizes. Using Laplace-Lagrange secular theory, we investigate the ability of stellar spindown to decouple PIAP systems. We find that strictly PIAP systems with regular spacings cannot undergo secular resonance crossings for the expected stellar $J_2$ evolution, and that we instead require the inner planet to migrate inward to undergo this resonance crossing. As a result, there is no inner edge to PIAP systems where systems will always cross a secular resonance and decouple the inner planet. Using expected $J_2$ evolution tracks from stellar evolution models, we find a diversity of expected resonance crossing times, highlighting the ability to test migration pathways and initial stellar obliquities using this framework.   
\end{abstract}

\keywords{Exoplanets (498), Exoplanet Astronomy(486), Natural Satellites (483): Exoplanet Evolution (491)}

\section{Introduction}\label{sec:intro}
Since the detection of the first transiting exoplanet \citep{Charbonneau2000}, transit surveys such as the \textit{Kepler} mission \citep{Borucki2010} and the Transiting Exoplanet Survey Satellite (\textit{TESS}; \citealt{Ricker2015}) have discovered the majority of the known populations of over 6,000 exoplanets.
Transiting planets are particularly useful in the study of exoplanetary systems, providing detailed information about planets' atmospheres \citep{Seahar2000, Feinstein2023}, orbital evolution \citep{Vissapragada2022}, and destruction \citep{Hon2025}.

One of the most striking discoveries by the \textit{Kepler} mission \citep{Borucki2010} was the prevalence of a particular type of architecture where the planets are regularly spaced and have similar sizes: the ``peas-in-a-pod'' (PIAP) systems \citep{Weiss2018, Weiss2023}.
These evenly spaced systems are one of the most common multi-planet architectures in the observed exoplanet census \citep{Howe2025}. 


In contrast with the PIAP systems, which appear typical in the exoplanet census, ultra-short-period planets (USPs) are an extreme class of planet with a relatively low absolute occurrence rate \citep{SanchisOjeda2014}, typically defined to be planets with an orbital period less than one day \citep{Sahu2006, Sanchis2013}. Although the USP boundary at $P\le 1$\,day was initially arbitrary \citep{Winn2018}, recent population statistics imply that the $P=1$\,day could be astrophysical, with distinct size and period ratio differences observed when compared to their non-USP counterparts \citep{Goyal2025}. Further, USPs tend to have larger mutual inclinations with their companions than non-USP systems \citep{Dai2018}, along with possessing an older age ($\gtrsim 1$\,Gyr) and having a higher occurrence rate around thick disk stars \citep{Tu2025}.

Because USP planets orbit interior to the magnetic truncation radius of protoplanetary disks \citep{Ghosh1979, bp1982, DAlessio1998}, traditional theories of planet formation struggle to form such planets at their observed radii. 
Moreover, USPs occupy extreme positions relative to the inner edges of the broader exoplanet population, reinforcing their status as architectural outliers \citep{Batygin2023}. As a result, recent work has focused on the ability of USP planets to migrate from larger initial orbital periods to their observed semi-major axes. 
Although several mechanisms have been proposed to explain the origin of USP planets, most invoke multi-body dynamics such as secular chaos or tidal interactions that excite and then damp eccentricity or obliquity and drive inward orbital evolution \citep{Schlaufman2010, Lee2017, Pu2019, Petrovich2019, Millholland2020}.

The statistical trend identified in \citet{Dai2018}, in which USP planets exhibit large mutual inclinations relative to adjacent planets, is generally robust. There also exist notable systems consistent with this general trend but with a distinctive geometry: a dynamically flat, high-multiplicity outer planetary system resembling a typical system of tightly packed planets, accompanied by a strongly misaligned inner planet \citep[e.g.,][]{Rodriguez2018, Quinn2019}.
A promising avenue of investigation is to more fully characterize the role that stellar spindown plays in mutual inclination excitation \citep{Spalding2016, Becker2020}. A time-varying $J_2$ of the host star causes the planetary orbits to precess, causing a variation of inclinations over time in a way that is very strongly orbital-radius-dependent \citep{LiGongjie2020, Chen2022}. A system can experience long-term mutual inclination excitation when the system undergoes a secular resonance crossing, which can project the innermost planet onto a large inclination and dynamically decouple it from the rest of the system \citep{Brefka2021,Faridani2025}, potentially rendering it undetectable to transit searches \citep{Faridani2025b}. Although this mechanism relied on a primordial misalignment between the host star's spin axis and the planetary axes at disk dissipation, recent work argues a significant fraction of outer protoplanetary disks are misaligned \citep{Biddle2025}, which could speculatively extend to the inner disk as well.

The intersection of PIAP and USP systems can then be examined to probe dynamical histories that may be unique to these extreme planets. When containing at least one other transiting companion, USP systems tend to have at least 3+ planets \citep{Adams2021}. The same work suggests that USP planets also have at least one other non-transiting companion. Multiplanet systems with USPs frequently show significant gaps between the USP and the PIAP outer planets (e.g. GJ-367, HD 3167, \citealt[]{Howe2025}) - which, by definition, are not a feature of PIAP systems. The prevalence of nearby (but not always transiting) companions in USP systems coupled with the dynamical tracers of heightened mutual inclinations \citep{Dai2018} and gaps lead one to posit that USPs may have originally started in a PIAP configuration with their neighbors and were later sculpted by a range of dynamical processes.

This work aims to link USP–PIAP system architectures to the evolution of their host stars by identifying which systems can undergo secular resonance crossings, and by assessing whether such crossings can account for the observed alignment properties of USP–PIAP systems.
This work is organized as follows: Section~\ref {sec:methods} outlines our assumed system setups (\S~\ref{sec:system_architecture}) along with the secular theory used to compute system evolution (\S~\ref{sec:secular_theory}) and our assumed stellar spindown prescriptions (\S~\ref{sec:J2}) used to compute its evolution; Section~\ref{sec:Simulations} contains a description of the simulated data and observed trends with system architectures; Section~\ref{sec:Discussion} discusses implications for general USP planet formation and compares our results with observed exoplanet system architectures. 
We then conclude in Section~\ref{sec:summary} with a succinct summary of our results.

\section{Methods} \label{sec:methods}

In this section, we describe the analytic machinery that we use to construct a suite of idealized compact multi-planet system architectures motivated by the overlap between the PIAP sample and the USP planet population, and define parameters to describe system architecture deviations from strictly PIAP configurations. 
For each modeled architecture, we describe how to compute the system’s secular nodal precession evolution using Laplace–Lagrange secular theory to identify when pairs of nodal eigenfrequencies evolve as the host star spins down and its rotational quadrupole moment, $J_2$, decreases with time. 

\subsection{System Architecture}\label{sec:system_architecture}
In this work, we investigate how deviations from an idealized PIAP system architecture, particularly those that introduce an ultra-short-period (USP) planet, affect a system’s susceptibility to secular resonance crossings.
To do this, we create a range of idealized system architectures based on a fiducial geometry. 
Our fiducial system has a central mass of $1\,M_\odot$ and a uniform planet mass of $10^{-5}\,M_\odot = 3.3\, M_\Earth$, corresponding to a $\approx 1.4$\,\RE planet using the mass-radius relationship of \cite{ChenKipping2017}, analogous to a typical super-Earth \citep{Fulton2017}.

In each system, we choose the period of the innermost planet ($P_\mathrm{inner}$) and then assign the orbital periods of additional planets by following an idealized finding of \cite{Weiss2018}, such that 
\begin{equation}
    \mathcal{P}= \frac{(P_{i+1}/P_i)}{(P_{i}/P_{i-1})}=1.
\end{equation} 
Although \cite{Weiss2018} found a distribution of $\mathcal{P}$ values ($1.03\pm 0.27$), we adopt a fiducial value of $\mathcal{P}=1$ to represent the ideal PIAP systems, which are equally spaced in $\log(P)$.

\begin{figure}[t]\label{fig:PIAP_comp}
    \centering 
\includegraphics[width=0.98\linewidth]{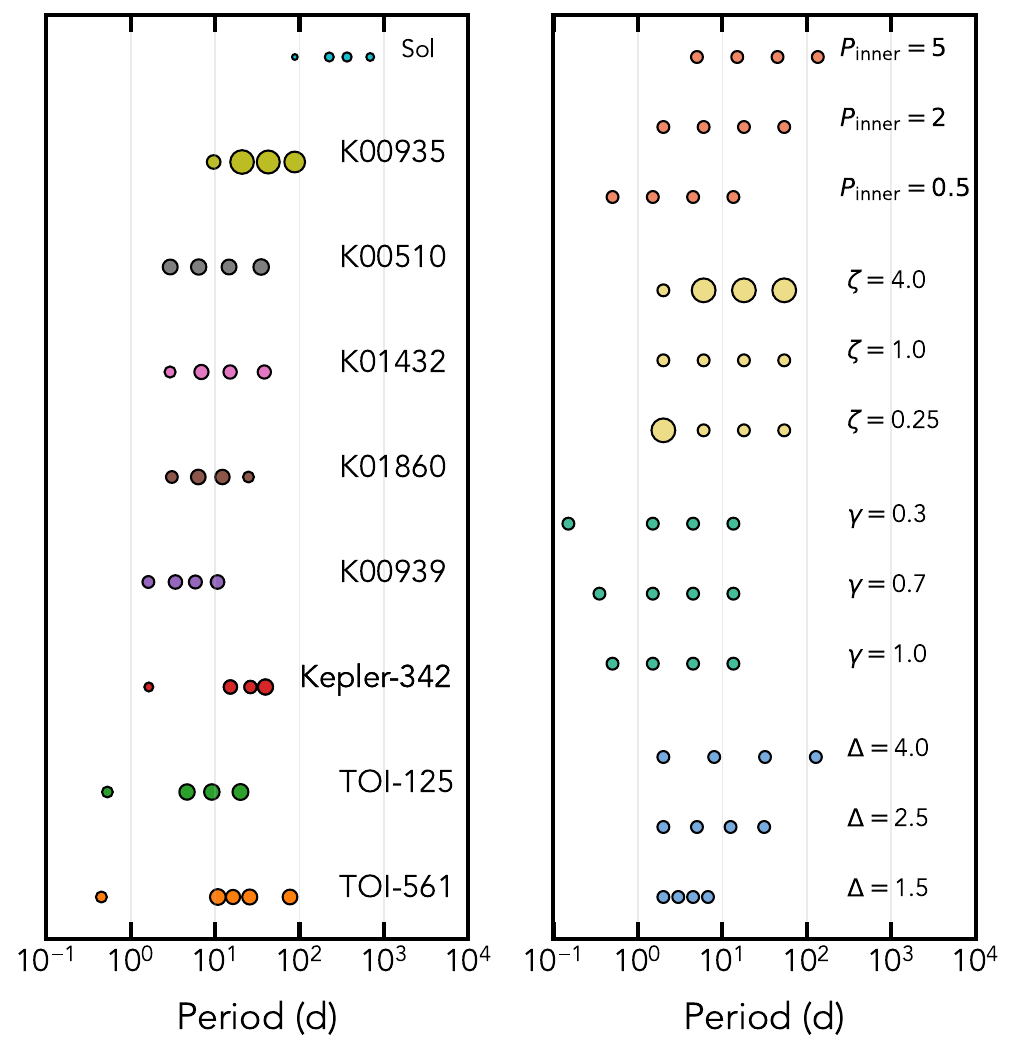}
    \caption{Left Panel: comparison between the solar system, typical Peas-in-a-Pod (PIAP) systems, and USP-hosting systems. Shown are USP-hosting systems such as TOI-125 \citep{Nielsen2020} and TOI-561 \citep{Piotto2024}; \textit{Kepler}-342 - a non-USP system with an inner gap \citep{Morton2016}; PIAP systems from the California Kepler Survey (identifier begins with a "K0"; \citealt{Petigura2017,Weiss2018}); and the solar system. These systems were chosen for illustrative purposes to highlight the trends seen in USP and PIAP systems. Right Panel: display of the different parameters varied in the system architecture in this work.}
    \label{fig:schematic}
\end{figure}

Interestingly, many of the planetary systems observed to host an USP planet are not observed in a PIAP configuration (see Figure~\ref{fig:PIAP_comp}), both in terms of planet spacing and planet size, with gaps typically seen between the USP and the rest of the observed planets. Further, some USP system show that the inner planet is smaller than their observed companions (e.g. \citealt{Quinn2019, Piotto2024}), potentially due to formation or evolutionary processes.

To create idealized system geometries, we define three geometric factors: 
\begin{itemize}
    \item \textbf{Period spacing ($\Delta$):}  
    After selecting the innermost period $P_{\mathrm{inner}}$, we construct the outer planets using a multiplicative spacing,
    $P_{i+1} = \Delta\, P_i,$
    where $\Delta$ ranges from 1.5 to 4, motivated by \textit{Kepler} period spacings \citep{Li2025} and consistent with the approximately uniform spacing in $\log P$ found by \cite{Weiss2018} and expanded on in \cite{Weiss2023}.  
    \item \textbf{Inner-period scaling ($\gamma$):}  
    We vary the orbital period of the innermost planet by a multiplicative factor $\gamma$, such that
    $P_{\mathrm{inner}}' = \gamma P_{\mathrm{inner}}$. 
    \item \textbf{Planet mass hierarchy ($\zeta$):}  
    We vary the relative masses of the planets using
    $\zeta = M_{\mathrm{outer}} / M_{\mathrm{inner}}$. In the case where we choose to invert the mass hierarchy ($\zeta <1$), we instead increase the mass of the inner planet so that $M_\mathrm{inner} > M_\mathrm{outer}=3.33$\,\ME. 
      
\end{itemize}

The orbital period $P_{\mathrm{inner}}$ of the innermost planet serves as the baseline for constructing the outer planets. All additional architectures are generated by applying the period-scaling factor $\gamma$ and the multiplicative spacing $\Delta$ relative to $P_{\mathrm{inner}}$. After choosing $P_\mathrm{inner}$, varying $\Delta$ allows for tuning of planet-planet coupling relative to stellar spindown, allowing for probing of $J_2$-dominated versus planet-planet-dominated regimes. $\gamma$ allows for the creation of a physical gap between the innermost and outer planets, seen in a variety of systems (e.g., TOI-125, TOI-561, and \textit{Kepler}-342 in Figure~\ref{fig:PIAP_comp}). Although PIAP systems typically contain planets with similar masses, we allow $\zeta$ to vary to reflect the variety of mass ratios seen in both USP and non-USP systems, allowing for both a smaller inner planet (e.g TOI-125, \citealt{Quinn2019}) and larger inner planet (e.g K2-266; \citealt{Rodriguez2018}). This relatively small set of parameters allows for synthetic systems to be generated that reflect the range of architectural diversity present in the current population of observed compact exoplanet systems.


\subsection{Secular Theory}\label{sec:secular_theory}
To investigate the mechanism of misalignment in the USP planet-hosting systems, we used the Laplace-Lagrange secular theory to model the evolution of the eigenfrequencies of the system in response to changing stellar $J_2$. For a system of $N$ planets around a single star, we can construct an $N\times N$ matrix ($B$) that describes the interactions between the planets that also account for stellar spindown, with coefficients defined as  \citep{Murray_and_dermott}

\begin{subequations}
\begin{equation}
    \begin{gathered}
        B_{jj} = -n_j \bigg{[} \frac{3}{2} J_2 (\frac{R_*}{a_j})^2 - \frac{27}{8} J_2^2 (\frac{R_*}{a_j})^4 
        \\ + \frac{1}{4} \sum_{k=1, \,k \neq j}^{N} \frac{M_k}{M_*+M_j } \alpha_{jk} \bar{\alpha}_{jk} b_{3/2}^{(1)}(\alpha_{jk}) \bigg{]},
    \end{gathered} 
    \end{equation} \vspace{0.1cm}
    \begin{equation}
        B_{jk} = + \frac{1}{4}  \frac{M_k}{M_*+M_j } n_j \alpha_{jk} \bar{\alpha}_{jk} b_{3/2}^{(1)}(\alpha_{jk})\,\,\,\,\,(j \neq k).
    \end{equation}
\end{subequations}
Here, $n_j = \sqrt{GM_* / a^3}$ is the mean motion of each planet, with $\alpha_{jk}$ and  $\bar{\alpha}_{jk}$ being defined as
\begin{subequations}
\begin{equation}
    \begin{gathered}
    \alpha_{jk} = \min(a_j, a_k) / \max(a_j, a_k),
    \end{gathered} 
    \end{equation}
    \begin{equation}
         \bar{\alpha}_{jk} = \min(1, a_j/a_k).
    \end{equation}
\end{subequations}
$b_{3/2}^{(1)}(\alpha_{jk})$ is a Laplace coefficient, calculated as
\begin{equation}
    b_{3/2}^{(1)} (\alpha) = \frac{1}{\pi} \int_{0}^{2\pi} \frac{\cos(\psi}{(1-2\alpha \cos(\psi) +\alpha^2)^{3/2}}\,d\psi.
\end{equation}

One can then find the secular nodal precession frequencies of a planetary system with $N$ planets by determining the eigenfrequencies of the $B$ matrix. When two precession frequencies in the system are commensurate, the two related modes become degenerate and the angular momentum deficit in the system may be redistributed. 
One way in which these precession frequencies may change with time is evolving stellar parameters such as the quadrupole moment $J_2$. It is important to note that the Laplace-Lagrange framework from \cite{Murray_and_dermott} does not account for tidal effects, which may be important for some of the USP planets we consider.

\subsection{$J_2$ Evolution} \label{sec:J2}
To model the spindown expected for main-sequence stars, we used stellar evolution models from \cite{Baraffe2015} and allowed the angular momentum to evolve using the methodology of \cite{Matt2015},
\begin{equation}
    \frac{d \Omega_*}{dt} = \frac{T}{I_*} - \frac{\Omega_*}{I_*} \frac{dI_*}{dt}\,,
\end{equation}
where $\Omega_*$ is the angular frequency of the star and $I_*$ is its moment of inertia. The torque from the stellar wind $T$ is further parameterized as 
\begin{equation}
    T= - T_0 \,\bigg(\frac{\tau_{cz}}{\tau_{cz \odot}}\bigg)^p \, \bigg(\frac{\Omega_*}{\Omega_\odot}\bigg)^{p+1}\,,
\end{equation}
with $\tau_{cz \odot}$ and $\Omega_\odot$ being the Solar convective turnover time and Solar angular frequency, respectively. $\tau_{cz}$ is the stellar convective turnover time, estimated using Eq. 36 of \cite{Cranmer2011}. We set $p=2$ to best match the typical scalings found in the literature ($T \propto \Omega_*^3 $, e.g \citealt{Chen2022} and \citealt{Faridani2023}). $T_0$ is the normalization factor defined as \citep{Matt2015, Matt2015Erratum}
\begin{equation}
    T_0 = 6.3 \times 10^{30}\,\mathrm{erg}\, \bigg( \frac{R_*}{R_\odot} \bigg)^{3.1} \bigg(\frac{M_*}{M_\odot} \bigg)^{0.5},
\end{equation}
where $R_\odot$ and $M_\odot$ refer to the Solar radius and mass. The variables $R_*$ and $M_*$ are the stellar radius and mass.

We used a forward Euler algorithm to evolve the angular momentum in time and calculated the corresponding $J_2$ evolution  \citep{Ward1976,Brefka2021}:
\begin{equation}
    J_2 = \frac{k_2}{3} \, \bigg( \frac{\Omega_*}{\Omega_{*,b}} \bigg)^2\,,
\end{equation}
assuming that all stellar deformation is due to rotation, with $k_2$ being the stellar Love number and $\Omega_{*,b}$ the stellar breakup frequency 
\begin{equation}
    \Omega_{*,b} \approx  \sqrt{\frac{GM_*}{R_*^3}}\,,
\end{equation}

with $G$ being the gravitational constant. We treat the initial frequency $\Omega_{*,0}$ as the rotation of the host star at disk dissipation, which we assume to be 10\,Myr. We take the apsidal motion constant from the stellar models in \cite{Claret2023}, and double it to obtain the Love number. As another point of comparison, we also utilize the stellar evolution tracks of \cite{Amard2019} (hereafter A19), which employ rotation self-consistently. For each system in \S~\ref{sec:timing}, we employ the closest median rotator track prescription given the system's mass and metallicity. For consistency and better observed agreement with solar values, with the A19 models we use $\Omega_* = 2\pi/P_\mathrm{rot}$ and the above definition of $\Omega_{*,b}$ for calculation of $J_2$ evolution.

\section{Secular Resonance Outcomes Across System Architectures}
\label{sec:Simulations}




To undergo the AMD redistribution described in \cite{Brefka2021}, a planetary system needs to cross a secular resonance during its lifetime. This will occur if multiple eigenfrequencies $\{\epsilon_i\}$, which describe the nodal precession frequencies in the system, approach the same value, in which case the related modes will become degenerate and allow the redistribution of AMD between modes.
The tangible result of such an exchange is changes in the observed orbital inclinations of the planets relative to each other and relative to the stellar equator \citep{Spalding2016}, which will affect the transit probabilities \citep{Faridani2025}.
This mechanism may either misalign inner planets from an outer transiting plane of planets \citep{Brefka2021, Chen2022}, or misalign outer planets from an inner transiting plane \citep{MacLean2025}, depending on the angular momentum hierarchy in the system.

In this section, rather than simulating the time evolution of individual systems through a secular resonance crossing, we determine how system geometry, parameterized by the architectural quantities defined in Section~\ref{sec:system_architecture}, governs whether a system is expected to cross a secular resonance during stellar spindown. 
The goal of this analysis is to determine whether PIAP systems that contain USP planets are expected to cross a secular resonance, or whether their typical orbital geometries make such crossings unlikely.

In the following analyses, we assume that tightly packed multi-planet systems (including those with USP planets) begin in nearly coplanar configurations following protoplanetary disk dispersal. Because passage through a secular resonance can redistribute orbital inclinations within a system, systems that do not encounter such a resonance are more likely to preserve their mutual coplanarity, even if the host star initially has an obliquity relative to the planetary orbital plane.

\begin{figure*}[t]
    \centering
\includegraphics[width=0.98\linewidth]{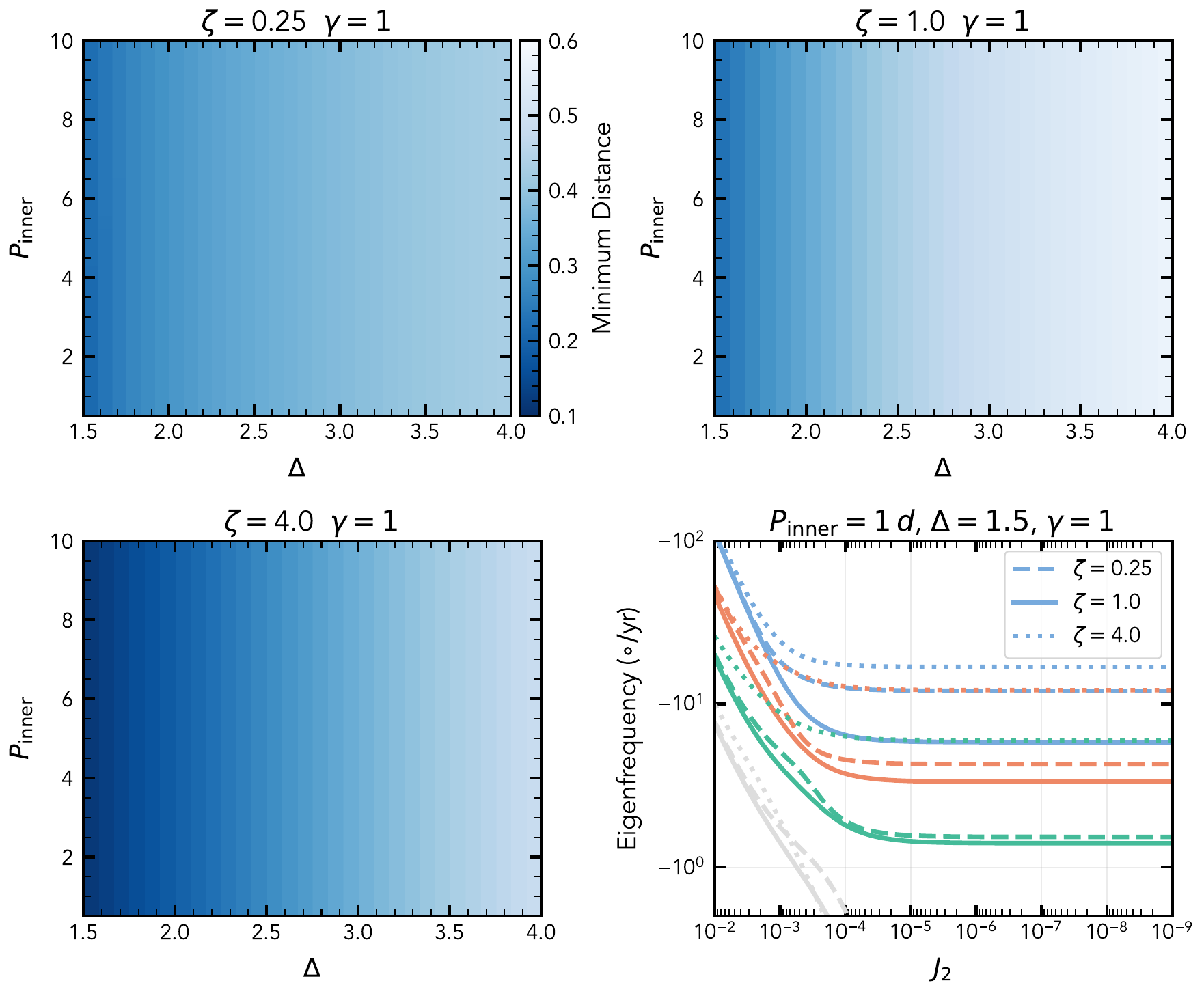}
    \caption{Minimum distances as a function of $P_\mathrm{inner}$ and $\Delta$ using the PIAP configuration ($\gamma=1$) across a variety of different outer planet masses. We let $\zeta=0.25,1,4$, corresponding to planets of super-Earth, mini-Neptune, and Neptune masses, respectively. We observe that the minimum distances are not dependent on the innermost period, and they tend to decrease as the outer planets' masses increase. None of the strictly PIAP systems is expected to cross secular resonances in our setup. Bottom right panel: example fundamental mode evolution for three different mass ratios and a specific choice of $P_\mathrm{inner}=1$\,d and  $\Delta=1.5$.}
    \label{fig:dp}
\end{figure*}

\subsection{Architectures Susceptible to Resonance Crossing}

Towards the goal of determining how likely different planetary system architectures are to pass through secular resonances, it is useful to define a metric to describe how close a system is to crossing a secular resonance at a particular $J_2$,
\begin{equation}
    D(\{\epsilon_i\}) = \min \bigg|\log_{10}|\epsilon_{i}| - \log_{10}|\epsilon_{j}|\bigg|_{i,j}\,\,.
\end{equation}
This metric allows us to evaluate the system across the full range of its $J_2$ evolution and identify the minimum separation from a secular resonance attained at any point in time. To determine which geometries allow secular resonance crossings, we run a set of parameter sweeps across our architecture parameters $P_{\rm{inner}},\ \Delta,\ \gamma$, and $\zeta$.

\begin{figure*}[t]
    \centering
    \includegraphics[width=0.95\linewidth]{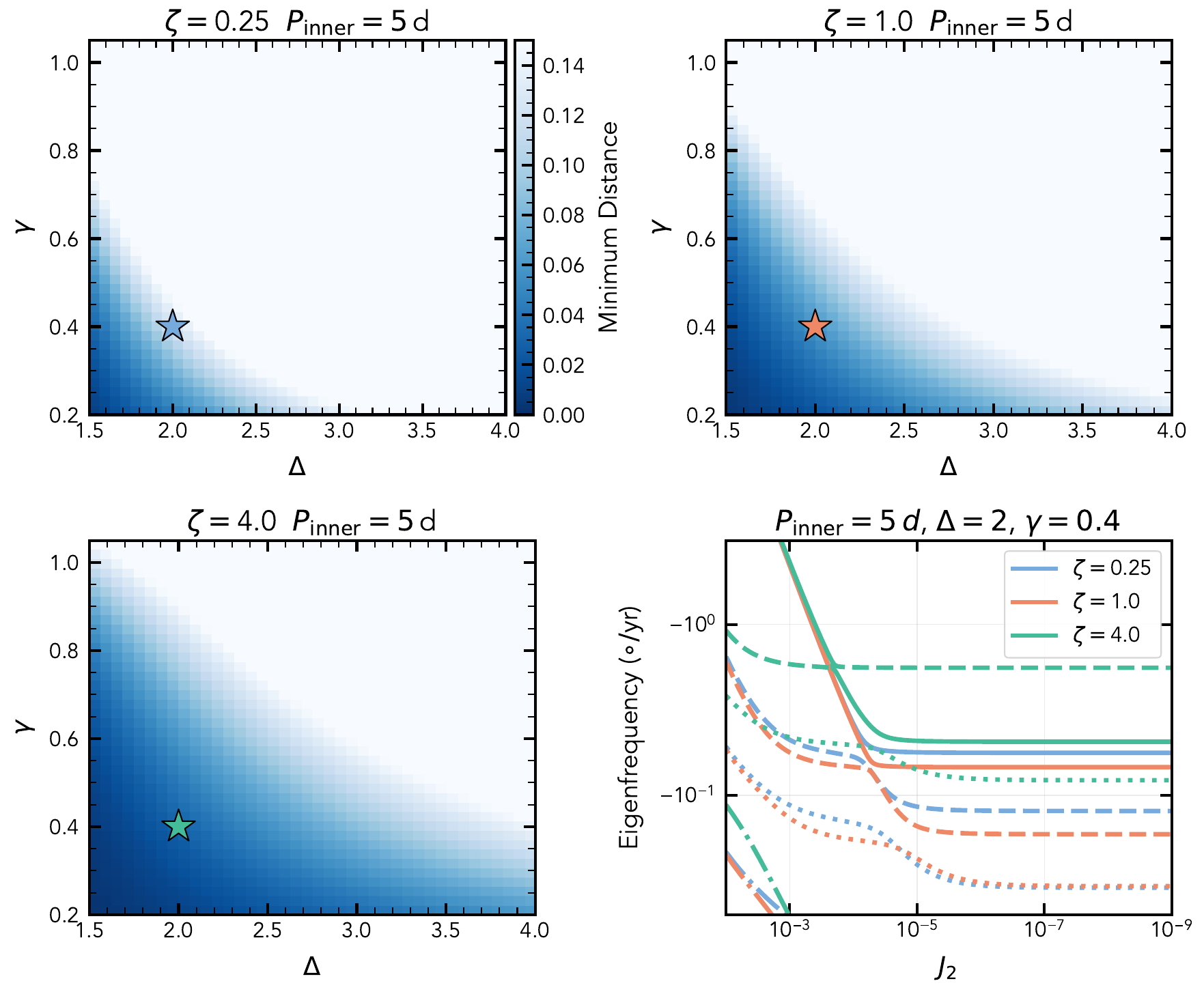}   
  \caption{Display of the proximity to a secular resonance as a function of $\Delta$ and $\gamma$. For each heat map, the initial inner period was fixed to 5\,d, and the zeta was varied between 0.25 and 4. Each heat map displays a star, corresponding to a specific example eigenfrequency evolution in the bottom right panel. We note that the upper colorbar bound of $D\le 0.15$ is arbitrary, and is used to highlight the shift in $D$ across different parameters. In the bottom right panel, three example evolutions are shown, each color-coded by the corresponding star on the respective panel.}
    \label{fig:gd}
\end{figure*}

%
%
%
%
In Figure~\ref{fig:dp}, we show the results of our first set of parameter sweeps across $\Delta$ and $P_{\rm{inner}}$ for $\gamma = 1$ and three discrete values of $\zeta$ ($\zeta = 0.25, 1.0, 4.0$; one per panel).  This chosen geometry generates synthetic systems that have period distributions consistent with the observed PIAP systems ($\gamma=1$) across a range of $P_\mathrm{inner}$ and $\Delta$, for three discrete values of $\zeta$ ($\zeta = 0.25, 1.0, 4.0$). 
The bottom right panel displays the eigenfrequency evolution for a sample of selected geometries under the $J_2$ spindown prescription described in Section \ref{sec:J2}. All geometries shown in this panel have equal spacings in log period ($\gamma = 1$). Each color denotes a different planet in the system and the line styles indicate different mass ratios ($\zeta = 0.25, 1.0, 4.0$). 
For this set of parameter sweeps, $\zeta = 1.0$ corresponds to a system with equal-mass planets, representing the prototypical PIAP configuration. In all examples shown, the eigenfrequencies do not cross, clearly indicating the absence of resonance crossing for PIAP systems, including those that include a USP planet.

While increasing the outer planets' masses (larger $\zeta$) decreases the closest approaches between planets in eigenfrequency space, these distances are independent of the innermost planet's period. 
Thus, we can conclude that even across a range of outer planet masses, strictly PIAP systems (as defined by their spacings) are not expected to cross nodal precession secular resonances as their host star spins down, even when such a system contains a USP planet. 
Further, we notice that mass ratios between the inner and outer planets do change the proximity to the secular resonance, with higher outer planet masses increasing the proximity to a secular resonance.

%
%
%
%
In Figure~\ref{fig:gd}, we show the results of our second set of parameter sweeps, where we vary $\Delta$ and $\gamma$ while holding $P_{\rm{inner}} = 5$ days and again varying $\zeta$ over three discrete values ($\zeta = 0.25, 1.0, 4.0$).  
Smaller values of $\gamma$ indicate that the innermost planet is more widely spaced from its nearest neighbor than the typical PIAP spacing.
This type of increased spacing is observed across a range of USP hosting systems (see the lower left corner of Figure \ref{fig:schematic}, where real systems Kepler-342, TOI-125, and K2-266 illustrate this geometry).
The first three panels of Figure~\ref{fig:gd} show that varying $\gamma$ and $\Delta$, while holding $\zeta$ and $P_{\rm inner}$ fixed, produces regions of parameter space in which a secular resonance is expected to be crossed, defined as where the minimum separation between the computed eigenfrequencies approaches zero.

There is a covariance between $\gamma$ and $\Delta$, such that more closely packed systems (low $\Delta$) combined with stronger decoupling of the innermost planet (low $\gamma$) yield closer proximity to secular resonances.
The lower-right panel of Figure~\ref{fig:gd} presents the evolution of the eigenfrequencies for three different values of $\zeta$, while keeping all other system parameters fixed ($P_{\rm inner} = 5$ days, $\Delta = 2.25$, $\gamma = 0.45$). 
Each set of planets is color-coded to match the corresponding parameter sweep, and each tested parameter set is indicated by a star of the same color in the first three panels.

We find that $\gamma$ and $\Delta$ are the strongest predictors of a secular resonance crossing as $J_2$ evolves. Systems that are more tightly packed (smaller $\Delta$) and have more weakly coupled inner planets (smaller $\gamma$) are more susceptible to secular resonances. The mass ratio $\zeta$ plays a secondary role, influencing the extent of parameter space that appears to permit a secular resonance.

%
%
%
%
%
%
%
%

\subsection{Observational Implications of Resonance-Crossing Timing}
\label{sec:timing}
Even if a system is predicted to encounter a nodal secular resonance based on the analysis in the previous sub-section, whether it actually does so depends on timing. Stellar spindown histories vary from star to star, and the $J_2$ value at which a system reaches resonance differs across system architectures. In this section, we will illustrate how this fact can be used to place constraints on individual system histories using three example systems which contain both a USP planet and a large number of outer, apparently coplanar planets: Kepler-80, K2-266, and TOI-125.

\begin{figure*}[htb!]
    \centering
    \includegraphics[width=0.99\linewidth]{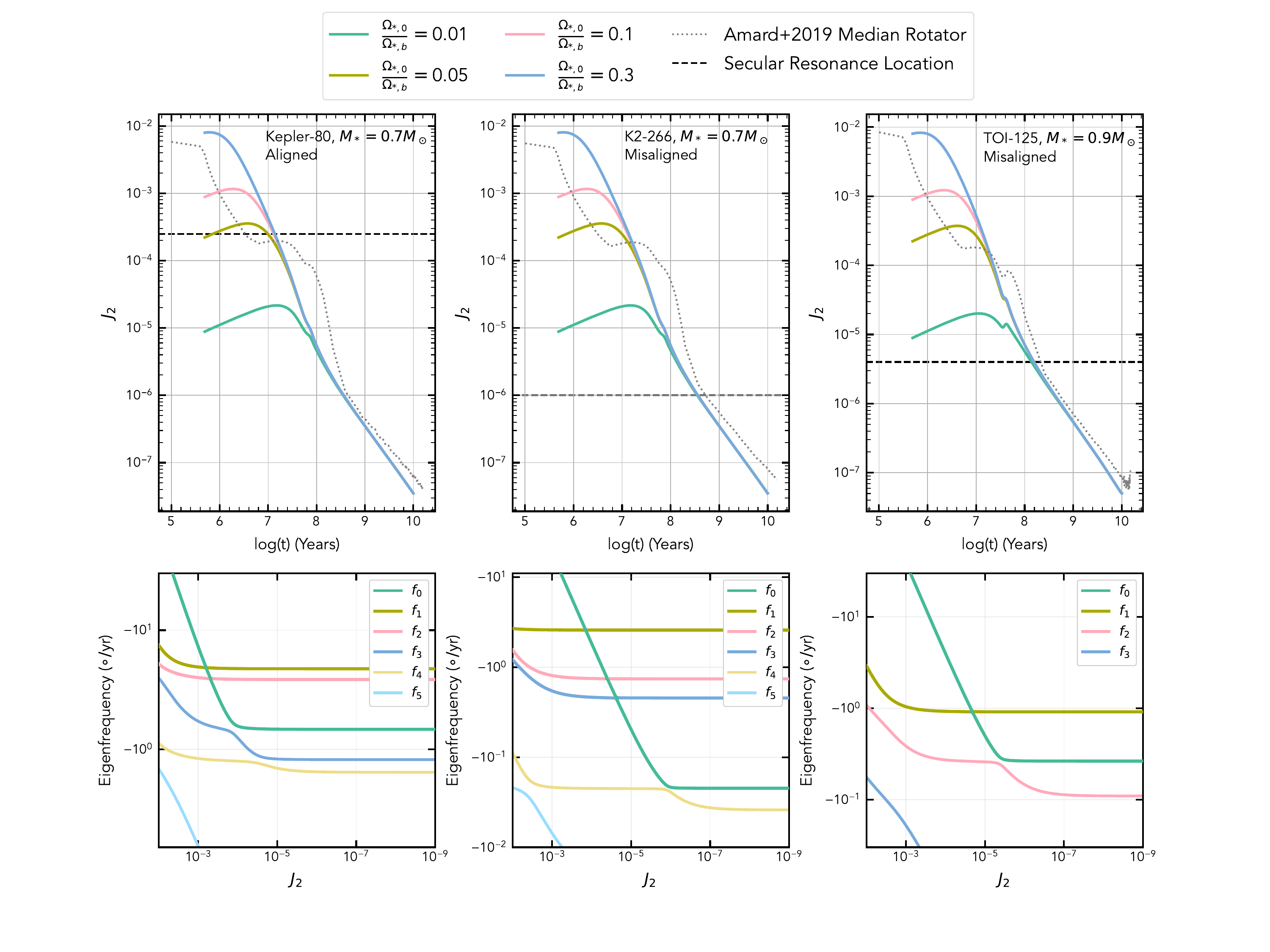}
    \caption{$J_2$ and eigenfrequency evolution for the \textit{Kepler}-80 system \citep{Macdonald2016}, K2-266 \cite{Rodriguez2018}, and TOI-125 \citep{Quinn2019} systems. Top panels: $J_2$ evolution for each system as described by \S~\ref{sec:J2} across a range of initial frequencies ($\Omega_*/\Omega_{b,*} = 0.01, 0.05, 0.1, 0.3$). The median rotator A19 models are shown as a dotted curve}. For each system, the line of the last secular resonance crossing is shown by a black dashed line. Bottom panels: eigenfrequency evolution as a function of $J_2$ for each system. The $J_2$ evolution was computed using the stellar models of \cite{Baraffe2015} and the angular momentum evolution prescription in \cite{Matt2015}.    \label{fig:USP_comp}

\end{figure*}

In the top panels of Figure~\ref{fig:USP_comp}, we present simulated $J_2$ evolution histories for three observed planetary systems: Kepler-80 (left), K2-266 (middle), and TOI-125 (right). For each system, we computed the $J_2$ evolution as a function of time as outlined in \S~\ref{sec:J2} along with the tracks of A19. Each panel shows models computed for different initial stellar spin frequencies, expressed as a fraction of the breakup rate. In the bottom panels of Figure~\ref{fig:USP_comp}, we show the eigenfrequency evolutions as a function of $J_2$, given the known planets and/or planet candidates in each system. 
All three of these systems are expected to undergo a secular resonance crossing at different values of stellar $J_2$. 
For each system, we identify the time of the most recent resonance crossing from the bottom panel and mark it as a dashed line in the corresponding top panel, indicating the system age at which the crossing occurs.

\textbf{Kepler-80}. In the left panels, we display the evolution of the Kepler-80 system, which consists of a system of six transiting planets \citep{Macdonald2016, Shallue2018}. Due to the aligned USP planet (\textit{Kepler}-80\,f) only possessing a mass upper bound \citep{Ofir2025}, we adopt the mass-radius relation from \cite{ChenKipping2017} as used by \cite{Louie2018} in the Terran regime, yielding an estimated mass of $1.92\,$\ME. The bottom center panel displays the evolution of the eigenfrequencies of the system, with clean crossings being observed near $J_2\sim 10^{-3}$ and $J_2 \sim 2\cdot 10^{-4}$. The top left panel shows the expected evolution of the $J_2$ of the $\approx 0.7\,M_\odot$ host star across a range of initial rotation frequencies, with the final secular resonance crossing ($J_2\sim 2\cdot 10^{-4}$) being shown as a horizontal dashed line. We observe that for our system to cross at least this final resonance, it requires $\tau \lesssim 20$\,Myr, which is quite rapid ($\tau \lesssim 5$\,Myr for the A19 models). This means for the Kepler-80 system to cross a secular resonance that could have misaligned the innermost planet, the initial $\Omega_{*,0}/ \Omega_{*,b} \gtrsim 0.05$ and the system must have assembled by 20 (5 for A19) Myr or so in its current configuration. 
The outer five planets of the Kepler-80 system appear to be in a chain of three-body mean-motion resonances \citep{MacDonald2021, Weisserman2023}, suggesting that the entire system migrated in a disk-driven way. Under the assumption that the inner planet in the system has to undergo some subsequent, post-disk dynamical mechanism \citep[e.g.][]{Petrovich2019} to take it to a final USP orbit, to allow this evolution to occur before the $J_2$ resonance would be crossed could be a significant timing challenge. 

\textbf{K2-266}. The center panels highlight the K2-266 system, which consists of a system of four confirmed transiting planets and two candidates \citep{Rodriguez2018}. In contrast to the \textit{Kepler}-80 system, the USP of the K2-266 system is very misaligned, a fact which has previously been attributed to potential early secular resonance crossings \citep{Becker2020}. 
Our analysis using only the four confirmed planets and their measured physical properties shows an expected secular resonance crossing of $J_2 \sim 6.2\cdot 10^{-6}$, which for a star of this mass would likely correspond to a stellar lifetime of just shy of 300\,Myr (or 600\,Myr for the A19 models). As a result of how late this secular resonance would be expected to occur, this system has a generous amount of time to reach its current geometry.

\textbf{TOI-125.} The rightmost panels of Figure \ref{fig:USP_comp} display the TOI-125 system, which hosts three candidate planets and a candidate USP planet \citep{Quinn2019, Nielsen2020}. Whereas \cite{Brefka2021} used the predicted masses reported in \cite{Quinn2019} for the TOI-125 system, we utilize the radial velocity-determined mass constraints from \cite{Nielsen2020} for the three outer mini-Neptunes. For the candidate misaligned USP for which a securely measured mass does not exist, we use the predicted mass of $2.65$\,\ME reported in \cite{Quinn2019}. Similar to the K2-266 system, the last resonance crossing is near 160\,Myr (250\,Myr for A19), leaving reasonable constraints on migration timescales.

\section{Discussion}\label{sec:Discussion}
Our results show that strictly peas-in-a-pod architectures ($\gamma=1$, $\zeta$ = 1) do not generically undergo nodal secular resonance crossings during stellar spindown, even when the innermost planet is a USP. 
Secular resonance crossings and the resulting dynamical decoupling of the inner planet become likely only when the innermost planet is weakly coupled to the outer system (small $\gamma$), which occurs in systems with an inner gap between the USP planet and the rest of the planets seen in the system. 

In the case that an otherwise PIAP system with a decoupled inner planet does have a geometry susceptible to resonance crossing, whether a USP ends up aligned or misaligned relative to the outer planet of PIAP planets depends primarily on timing: the inner planet must migrate into its USP orbit before the host star’s $J_2$ decays past the resonance.

If secular resonance crossing is the main culprit of mutual inclination excitation between inner planets and outer planes of planets in compact multi-planet systems (geometries seen in several observed systems including K2-266 and TOI-125), USP planets that fail to migrate into their USP orbits fast enough will fail to reach the $\gamma$ needed to cross a secular resonance, leaving them aligned with the remainder of the system. 

Under the assumption that nodal secular resonance crossings are the primary mechanism driving mutual inclination excitation in USP-hosting systems, we propose the following general evolutionary pathway that naturally explains both aligned and misaligned USP architectures:
\begin{enumerate}
    \item The planets begin in a coplanar PIAP configuration from forming in the protoplanetary disk, with a roughly constant $\Delta$ \citep[from, perhaps, energy equipartition enforced during formation;][]{Adams2019, Adams2020} and $\gamma=1$ and some non-zero primordial misalignment between the stellar spin axis and the plane of the planets' orbits at disk dissipation. 
    \item Secular chaos, tidal forces, or another mechanism operates to shrink the orbit of the USP, decreasing its value of $\gamma$.
    \item Simultaneously, the star is spinning down, decreasing its $J_2$ value. As shown in Section~\ref{sec:Simulations}, strictly PIAP systems will not typically cross secular resonances regardless of $J_2$ values. Thus, the planet needs to sufficiently decrease its $\gamma$ fast enough to ``catch'' the $J_2$ secular resonance crossing. Thus, computing the expected time of crossing for secular resonances and observing the architectures of the resulting planetary systems can directly probe migration mechanisms.
\end{enumerate}







\subsection{The Inner Edges of Peas-in-a-Pod Systems}


In this paper, rather than examining the exact geometries of known PIAP systems, we use a set of parameterizations of system architecture to create generalized systems and examine which general system properties are expected to be crossed at particular resonances. This allows us to identify that the parameter space populated by PIAP systems (defined by $\zeta=1$, $\gamma=1$) is unlikely to have passed through nodal precession secular resonances in the past, which would have potentially excited mutual planetary inclinations and moved some planets out of the transiting plane. 
This remains true even in systems where the PIAP system contains an evenly spaced USP planet (as shown in Figure \ref{fig:dp}).
As a result, we would expect that if PIAP systems had inner USP planets, they would not have been perturbed out of the transiting plane due to this mechanism.

One might hypothesize that if a PIAP system migrates close enough to its host star, $J_2$ evolution would dominate for the inner planet and force a secular resonance crossing, resulting in an ``inner edge" to the PIAP systems. Instead, we find a gap is required ($\gamma<1$) to undergo a nodal secular resonance crossing, which can happen even for a system without a USP. Thus, the inner edge to PIAP systems is likely not set by this mechanism of secular resonance crossing, but by the migration mechanism causing the gap.


\subsection{Observational Signatures of Nodal Precession Secular Resonance Crossings in the Exoplanet Sample}
\citet{Howe2025} and \citet{Howe2026} identify two main, at times overlapping, architectures classes in the closely-spaced exoplanet sample: peas-in-a-pod and gapped systems. In our formulation, systems where $\gamma <<1$ are inner-gap systems (see Figure 1 of \citealt{Howe2025}). 
Our results demonstrate an expected difference in orbital properties between PIAP systems and inner-gap systems: the latter are expected to have larger dispersions in orbital inclinations as a result of nodal precession secular resonances. 

As we discuss in Section \ref{sec:timing}, the timing at which a secular resonance occurs and the time at which the USP planet reaches its observed orbital period together will determine whether the secular resonance occurs and whether mutual inclination will be excited between the USP planet and the rest of the planets in the system. 
The observed exoplanet distribution will be sculpted by the interplay between these two processes \citep[e.g.,][]{Lam2024}, meaning that the inclination distribution of short-period planets contains some information about how often the nodal precession secular resonance onsets. Recent observational results \citep{Schmidt2024, Tu2025} suggest that USP planets attain their final orbits later than other geometries of exoplanet system, meaning that many such systems would be expected to miss the secular resonance. 

For individual systems, their observed geometries can be used as suggestions, if not direct evidence, that the system assembled early enough to be affected by the secular resonance. For example, for the TOI-125 system analyzed in \cite{Brefka2021}, $\Delta \approx 2$ from the outer 3 planets, implying $\gamma \approx 0.2$. Further, the outer planets are more massive than the USP in the TOI-125 system, with $\zeta \approx 3.4$. From Figure~\ref{fig:gd}, one can see that TOI-125 would be expected to cross a secular resonance during its host star's spindown, agreeing with the previous result of \cite{Brefka2021}. 

We note that for the population of USP planets, low or zero observed current-day eccentricities are not surprising even in the context of hypotheses that involve excited orbital eccentricity such as that proposed by \citet{Petrovich2019} and the more recent arrivals in their observed orbits proposed by \citet{Schmidt2024} and \citet{Tu2025}.
To demonstrate this, we can compute the circularization timescale $\tau_\mathrm{circ}$ \citep{Jackson2008}:
    \begin{equation}
        \tau_\mathrm{circ} \equiv \bigg{[} \frac{1}{a} \frac{da}{dt} \bigg{]}^{-1} = \frac{2}{63} (GM_*^3)^{-1/2} \frac{Q_p M_p}{R_p^5} \frac{a^{13/2}}{e^2}. 
    \end{equation}
For these close-in rocky planets, the circularization timescales are expected to be $\lesssim 10^5-10^6$\,yr using an assumed $Q_p \sim 10^2$ \citep[appropriate for Earth-like or super-Earth-like planets;][]{Goldreich1966}. For this reason, while mutual inclination with nearby planets can be used as a dynamical tracer to infer a USP planet's history, orbital eccentricity cannot be used in the same way. 
At the same time, this effect could also lead to systematic misinterpretation of USP planet–hosting systems \citep[for example, by causing multi-planet systems to often appear as though they contain only a single USP planet from certain lines of sight;][]{Ballard2016, Moriarty2016}.

One might worry that the inclination signature might also be eroded over time due to stellar obliquity damping, or, instead that a primordially misaligned disk might damp its innermost planet's obliquity more than the outer companions, artificially creating a heightened mutual inclination. To ensure the robustness of inclination as a tracer of dynamical history, we use the convective envelope timescale presented in \cite{Albrecht2012}:
\begin{equation}
    \tau = 10^{10}\,\mathrm{yr}\,\bigg(\frac{M_p}{M_*}\bigg)^{-2} \bigg(\frac{a/R_*}{40}\bigg)^{6}.
\end{equation}
Using the parameters in our simulated exoplanet architectures, along with a value of $a/R_*=2$, we arrive at a damping timescale of $\sim 10^{12}$ years. Thus, USP systems would be expected to retain their initial stellar obliquities.
\subsection{Future Work}
It should be noted that planets not crossing secular will indeed still undergo inclination variations as a result of stellar spindown \citep[e.g.,][]{Batygin2016}; however, this means that the mutual inclinations will generally be time-variable but centered around the same plane, resulting in a system where all planets might be seen in transit at some times and not at others due to large mutual oscillations in inclination between all planets. 
Instead, if the system migrates fast enough to reach its final orbits before the star passes the $J_2$ value corresponding to a secular resonance crossing, the innermost planet can be projected out of the plane of the other planets, decoupling it dynamically and imbuing it with a relatively time-invariant mutual inclination with the rest of the planets. 
If secular resonance crossing is the dominant influence shaping the inclination distribution of USPs, then analyses of USP can yield insights into planetary migration timescales and interiors, primordial stellar obliquities, and initial rotational frequencies of their host stars.  

We also note that in this work we identified the regimes in which nodal precession secular resonances are expected to occur, but we did not model the detailed dynamical evolution that follows once such a resonance is crossed. Previous studies \citep[e.g.,][]{Brefka2021, Chen2022, Faridani2023} have investigated this evolution using numerical simulations for specific system geometries. Future work could extend similar analyses to other systems in which secular resonance crossings are expected to play an important role in shaping the system’s dynamical history.

Finally, there are additional caveats in using our framework to infer system histories. Because our results are highly sensitive to physical parameters (particularly planetary mass ratios between adjacent planets, or a missing planet in a gap that was not detected), observational uncertainties in measured planet properties may significantly affect the inferred outcomes for individual systems. Our result also neglects the expected evolution of the planet's mass over the first few hundred Myr, with photoevaporation and core-powered mass loss likely playing a key role in driving atmospheric escape \citep{OwenWu2017, Ginzburg2018}. Consequently, determining whether secular resonance crossings are expected to be important for a given system geometry requires well-constrained planetary parameters and evolutionary models.

\section{Summary}\label{sec:summary}
In this work, we use Laplace-Lagrange secular theory to test the inner edge of the PIAP systems. We find that under expected $J_2$ values from stellar evolution, strictly PIAP systems do not cross secular resonances and thus remain coupled in orbital inclination. Instead of the innermost period determining the crossing of a secular resonance, we require a gap between the innermost planet and the second planet to cross the secular resonance ($\gamma <1$), with period spacing ($\Delta$) and mass ratios ($\zeta$) also playing a role.

In light of this, we identify the need for migration to occur to decouple the innermost planet from a secular resonance crossing, highlighting the need for a migration pathway to invoke this mechanism. Further, due to the variety in expected resonance crossing times for observed USP systems, we show that some systems may not cross secular resonances due to early resonance crossing times.

\begin{acknowledgments}
AD gratefully acknowledges the generous support of the \textit{Peter Livingston Scholars Program}, whose contributions to undergraduate research have played a key role in the development of this work.
This material is based upon work supported
by the National Science Foundation Graduate Research Fellowship under Grant No. DGE-2140743. Any opinion, findings, and conclusions or recommendations expressed in this material are
those of the authors(s) and do not necessarily reflect the views of the National Science
Foundation.
\end{acknowledgments}
\software{
\texttt{matplotlib} \citep{Hunter2007},
\texttt{numpy} \citep{harris2020array},
\texttt{scipy} \citep{2020SciPy-NMeth},
\texttt{pandas} \citep{reback2020pandas}.
}

\bibliographystyle{aasjournalmod}
\bibliography{biblio.bib}

\end{document}